\documentclass[conference]{IEEEtran}
\IEEEoverridecommandlockouts
\usepackage{cite}
\usepackage{amsmath,amssymb,amsfonts}
\usepackage{algorithmic}
\usepackage{graphicx}
\usepackage{textcomp}
\usepackage{xcolor}
\usepackage{graphics}
\usepackage{graphicx,amsmath,epstopdf,caption}
\usepackage{flushend}
\usepackage{multirow}
\usepackage{subcaption}
\usepackage{calc}
\newlength{\subfigurewidth}

\def \sys {\textit{Crescendo}}
\def\BibTeX{{\rm B\kern-.05em{\sc i\kern-.025em b}\kern-.08em
    T\kern-.1667em\lower.7ex\hbox{E}\kern-.125emX}}
\begin{document}
\graphicspath{ {./}}
\title{\sys{}: An Infrastructure-free Ubiquitous Cellular Network-based Localization System}
\author{\IEEEauthorblockN{Rizanne Elbakly}
\IEEEauthorblockA{\textit{Wireless Research Center} \\
\textit{Egypt-Japan Univ. of Science and Technology (E-JUST)}\\
Alexandria, Egypt \\
rizanne.elbakly@ejust.edu.eg}
\and
\IEEEauthorblockN{Moustafa Youssef}
\IEEEauthorblockA{\textit{Dept. of Computer and Systems Engineering} \\
\textit{Faculty of Engineering, Alexandria University}\\
Alexandria, Egypt \\
moustafa@alex.edu.eg}
}

\maketitle
\begin{abstract}
A ubiquitous outdoor localization system that is easy to deploy and works  equally well for all mobile devices is highly-desirable. The GPS, despite its high accuracy, cannot be reliably used for this purpose since it is not available on low-end phones nor in areas with low satellite coverage. The application of classical fingerprinting approaches, on the other hand, is prohibited by excessive maintenance and deployment costs.

In this paper, we propose \sys{}, a cellular network-based outdoor localization system that does not require calibration or infrastructure support. \sys{} builds on techniques borrowed from computational geometry to estimate the user's location. Specifically, given the network cells heard by the mobile device it leverages the Voronoi diagram of the network sites to provide an initial ambiguity area and incrementally reduces this area by leveraging pairwise site comparisons and visible cell information.

Evaluation of \sys{} in both an urban and a rural area using real data shows median accuracies of 152m and 224m, respectively. This is an improvement over classical techniques by at least 18\% and 15\%, respectively.
\end{abstract}


\section{Introduction}

Nowadays, finding one's location outdoors is usually based on the  GPS \cite{misra1999special}. Although the GPS can work virtually anywhere around the world, it is not available on all mobile devices, can fail to find a location (e.g. when the signals to the satellites are obstructed inside tunnels or because of the urban canyon effect \cite{aly2013dejavu}), and consumes a lot of power. Therefore, other techniques of outdoor localization need to be developed to cover scenarios such as energy-efficient localization or emergency response tracking (E911), where all users regardless of their devices' capabilities need to be tracked.
To address that, a number of approaches have been recently proposed, including fingerprinting-based~\cite{ibrahim2012cellsense,ergen2014rssi}, sensor-based~\cite{ali2018senseio,wang2017woloc,wang2018learning,aly2013dejavu,alzantot2012uptime,aly2017accurate,aly2015lanequest,aly2016robust,aly2015semmatch,aly2014map++,constandache2010towards,youssef2010gac}, and cellular network-based~\cite{ibrahim2012cellsense,ibrahim2013enabling,chakraborty2015network,liu2015mobile,mohamed2014accurate,zhu2016city,huang2017experimental,paek2011energy, abdelaziz2015diversity}. In addition, due to the availability of both computation resources as well as huge datasets, neural networks and deep learning have been recently leveraged for localization~\cite{abbaswideep,rizk2018cellindeep,elbakly2018truestory,shokry2018deeploc,rizk2019effect}. According to the experimental study in~\cite{huang2017experimental} probabilistic cellular network fingerprinting in~\cite{ibrahim2012cellsense} provides the highest accuracy for 2G networks. However, systems such as~\cite{ibrahim2012cellsense,ergen2014rssi} require an onerous and time-consuming wardriving phase either to collect a cellular signal fingerprint or to build a model based on the road network and typical driver behavior.
Sensor-based systems use sensors available on high-end phones such as WiFi~\cite{wang2018learning,wang2017woloc} and inertial sensors~\cite{aly2013dejavu,constandache2010towards,youssef2010gac}. These techniques do not work for low-end phones. 

In contrast, cellular network-based techniques are by default available for all phones. The most basic approach is the Cell ID method~\cite{dufkova2008active}, where the location is estimated as the longitude and latitude of the strongest visible network cell. Despite its simplicity, the Cell ID method has a coarse-grained accuracy. Cellular fingerprinting approaches~\cite{ibrahim2012cellsense,chakraborty2015network} provide superior accuracies, yet require time-consuming data collection.
To completely avoid labeled dataset collection a propagation model is leveraged in~\cite{liu2015mobile} to generate the fingerprint. However, this significantly affects the accuracy of the system.
In this paper, we propose \sys{}: a ubiquitous cellular network-based localization technique that does not require neither calibration nor any additional support from the already existing infrastructure (the cellular network). \sys{} starts its operation by building the Voronoi diagram of the area of interest with network sites as seeds. Since distance is typically inversely proportional to the Received Signal Strength (RSS), the user is assumed to be closest to the strongest visible network site and can hence be placed in the Voronoi polygon of the strongest site. %
Thereafter, signal information of cellular network cells visible to the user's phone are used to incrementally narrow down this initial ambiguity region. Specifically, instead of using a propagation model or absolute RSS values as is commonly done in other approaches, the \textit{relative} RSS between each pair of heard cells/sites at the device is used to constrain the user's location and hence reduce her ambiguity region.
Additionally, available information about the sector covered by each visible cell is used to improve the accuracy of the location estimate.

We have implemented and tested \sys{} in urban and rural areas with different network densities. The results show that \sys{} can achieve a median accuracy of 152m and 224m in the urban and rural areas, respectively. This accuracy is an improvement over classical infrastructure-free techniques by at least 18\% and 15\% and comes with no calibration overhead.

The rest of this paper is organized as follows: in Section~\ref{sec:basic} we briefly give a background on the structure of the cellular network and the terminology used throughout the paper as well as give an overview of the basic idea of \sys{}. The full system architecture  and evaluation of the system are presented in Sections~\ref{sec:full} and~\ref{sec:eval}, respectively. Section~\ref{sec:conc} concludes the paper.

\begin{figure}[!t]
	\centering
	\includegraphics[width=0.8\columnwidth]{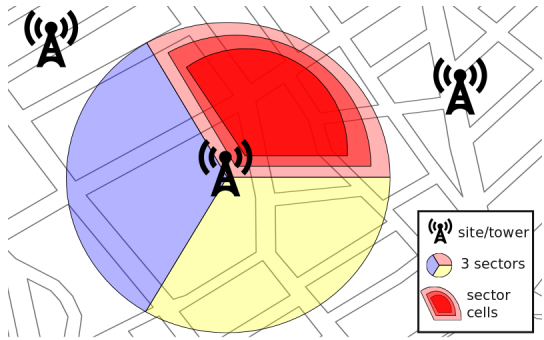}
	\caption{Basic cellular network architecture.}
	\label{fig:network}
\end{figure}

\section{\sys{} Basic Idea} \label{sec:basic}
\subsection{Background}
\label{sub:network_structure}

As shown in Fig.~\ref{fig:network} the cellular network consists of multiple  sites, which are distinct physical locations - also called \textit{towers}. Each tower covers an area around it divided into multiple physical \textit{sectors}. A sector represents a certain ``slice'' of the circle centered around the site, and is represented by two straight lines emanating from the site location at a specific inclination from north. Theoretically, the sector area can be considered extending to infinity. However, in practice due to signal attenuation it is finite as shown in Fig.~\ref{fig:network}. Multiple \textit{cells} operating at different frequencies cover the same sector. A mobile unit (MU), i.e. a phone, can detect up to seven different cells at the same time instant and is associated to only one of them. The different visible cells can be from the same site or from different sites, and can be even covering the same sector. 
Information about the RSS of visible cells is available at the provider-side as well as at the device-side.

\subsection{Basic Idea}
In this section we start by explaining the basic idea of the proposed algorithm and how it works under ``ideal'' conditions (Sections~\ref{sub:vor_constraints} and~\ref{sub:cell_info}). 
Here, we assume a hypothetical ideal environment and cellular network, where a) the propagation environment is ideal, i.e. a cell is visible only within the area defined by its sector. Therefore, if a MU detects a cell $c$, it has to be located within the area defined by its sector $r$. Users outside $r$ cannot hear any cells covering $r$; and b) a MU can detect up to seven cells but only one cell per physical site/tower, i.e. cells visible in a single scan are located at distinct sites. 

We \textbf{relax these assumptions} later in Section~\ref{sec:full}.

\begin{figure*}[htp]
	\setlength{\subfigurewidth}{.33\textwidth}
	\begin{subfigure}{\subfigurewidth+1em}
		\includegraphics[width=\subfigurewidth]{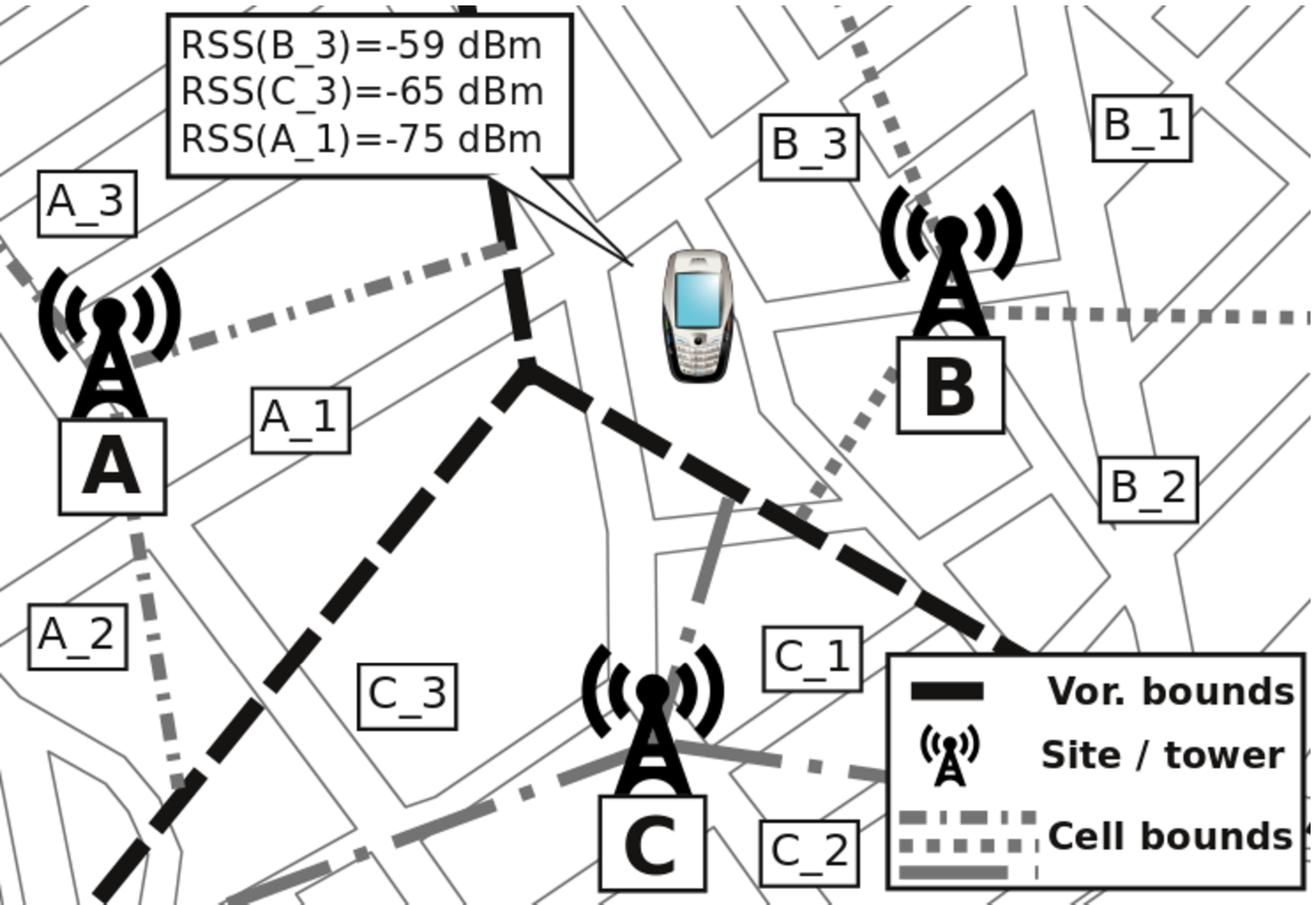}%
		\hfill
		\caption{Voronoi diagram with sites as seeds.}
		\label{fig:pairwiseConstraints_1}
	\end{subfigure}%
	\begin{subfigure}{\subfigurewidth+1em}
		\includegraphics[width=\subfigurewidth]{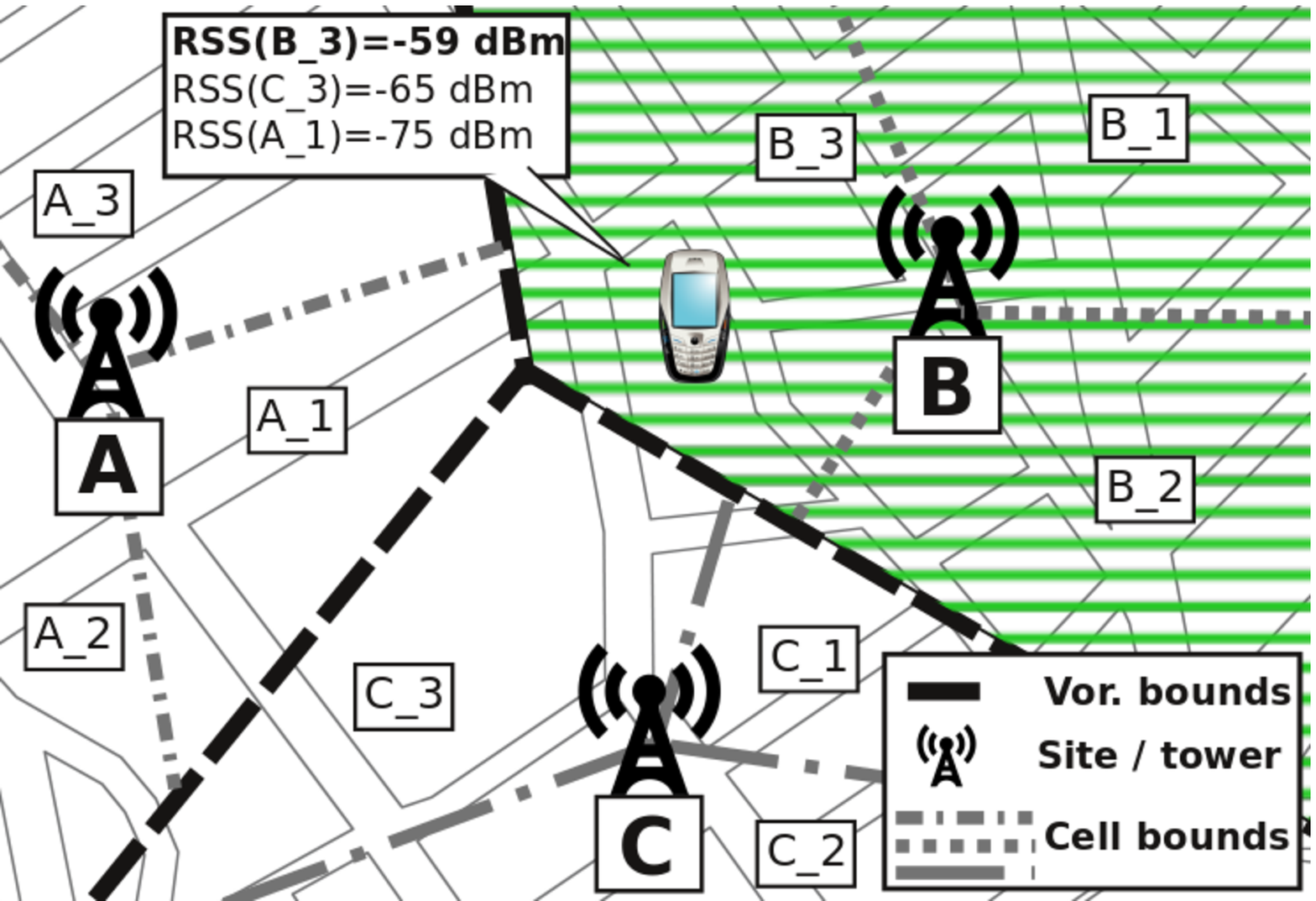}%
		\hfill
		\caption{Initial ambiguity area.}
		\label{fig:pairwiseConstraints_2}
	\end{subfigure}%
	\begin{subfigure}{\subfigurewidth+1em}
		\includegraphics[width=\subfigurewidth]{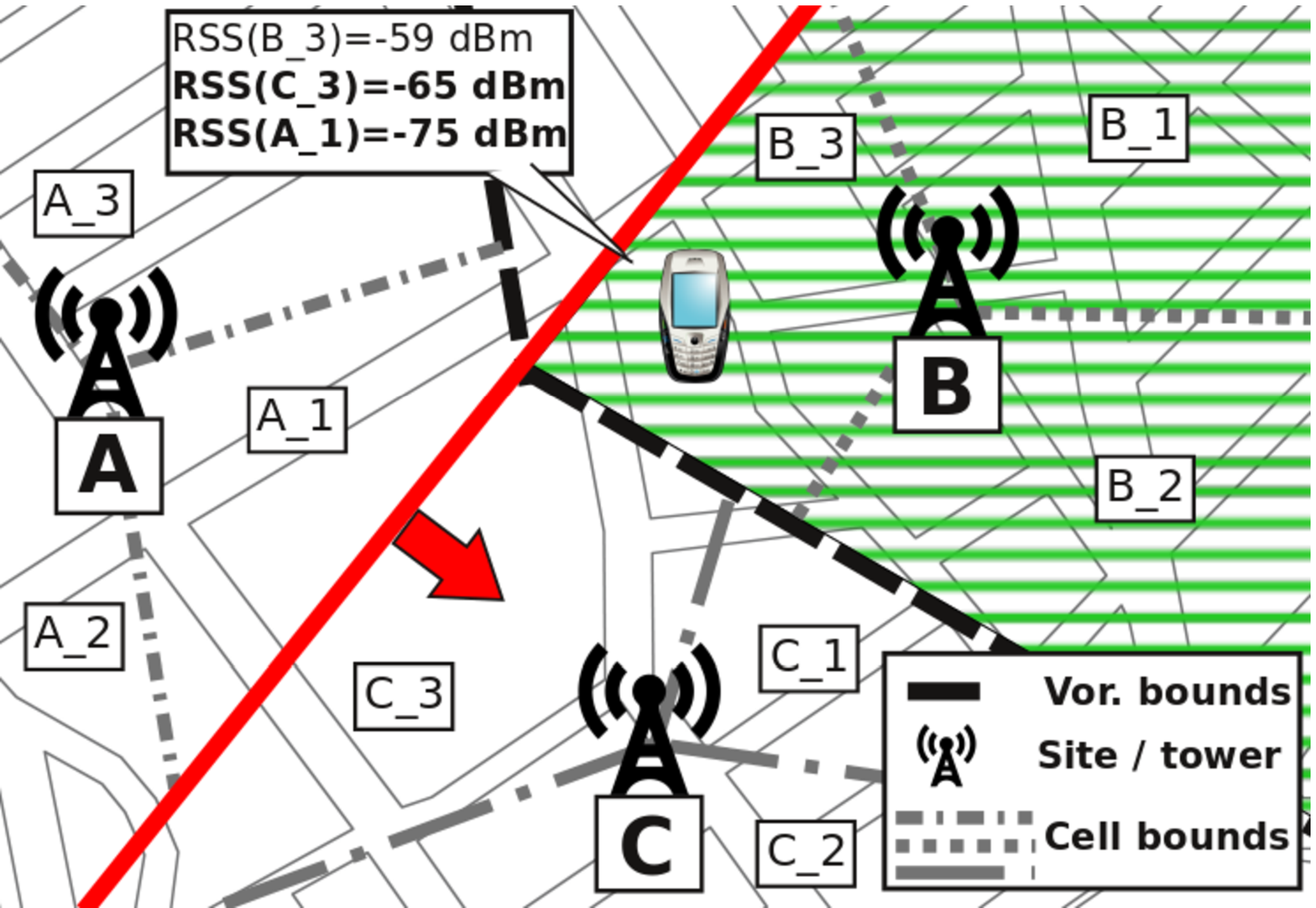}%
		\hfill
		\caption{Pairwise Site/Cell Constraint.}
		\label{fig:pairwiseConstraints_3}
	\end{subfigure}%
	\caption{Step 1: Initial location estimation based on Voronoi diagram of strongest site and pairwise site comparisons.}
\end{figure*}

\begin{figure*}[htp]
	\setlength{\subfigurewidth}{.33\textwidth}
	\begin{subfigure}{\subfigurewidth+1em}
		\includegraphics[width=\subfigurewidth]{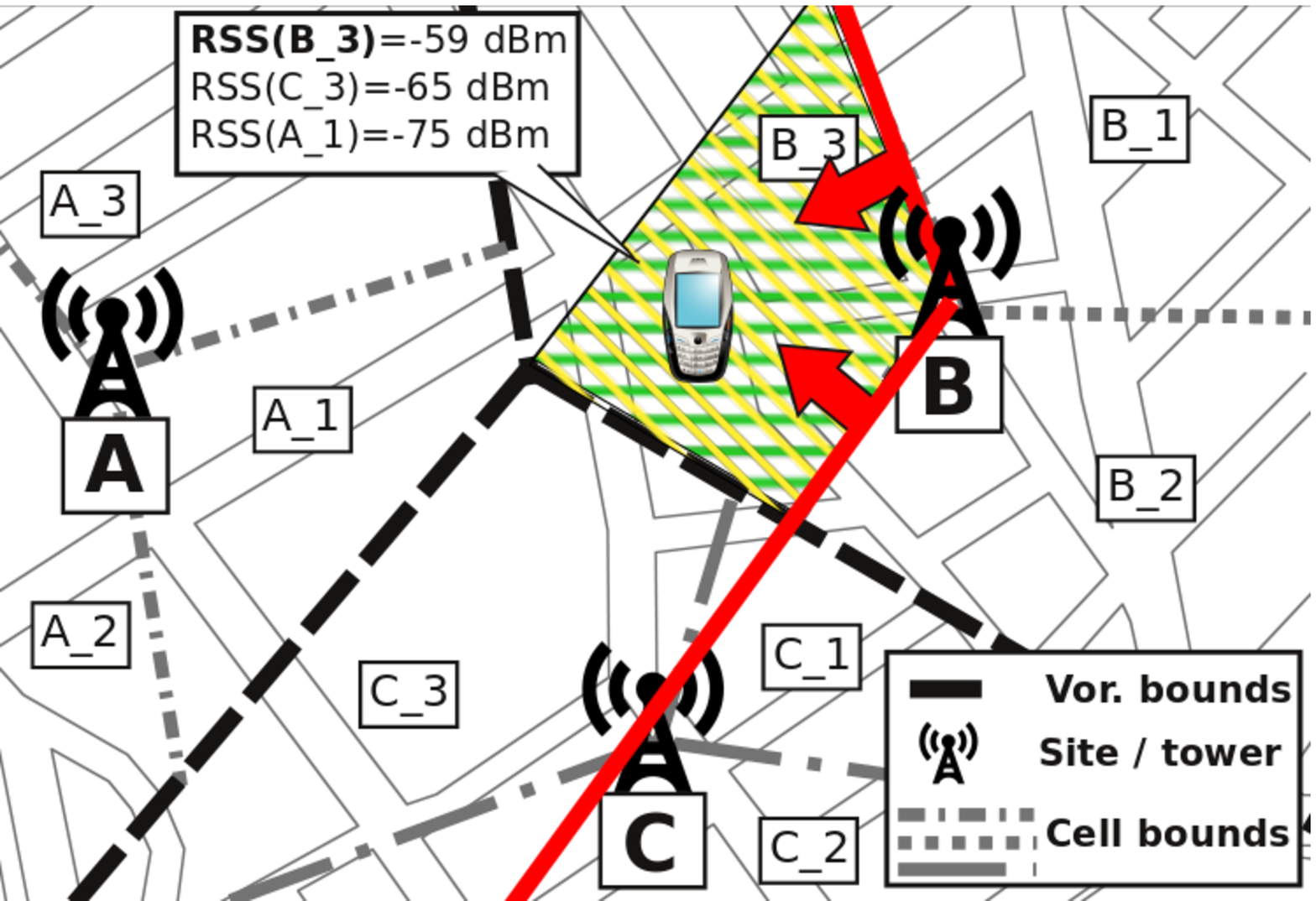}%
		\hfill
		\caption{Cell $B\_3$ contains the location.}
		\label{fig:cellConstraints_1}
	\end{subfigure}%
	\begin{subfigure}{\subfigurewidth+1em}
		\includegraphics[width=\subfigurewidth]{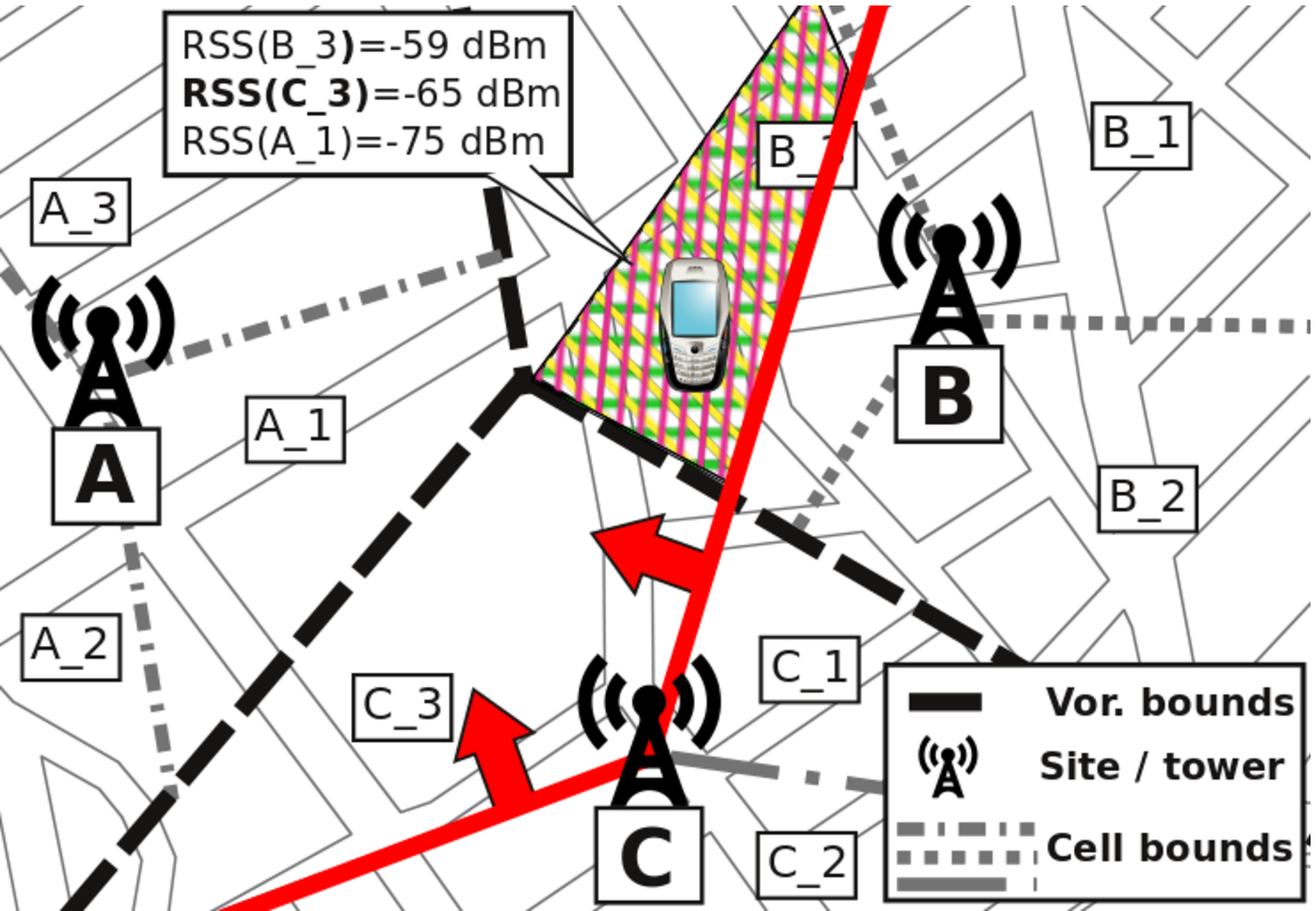}%
		\hfill
		\caption{Cell $C\_3$ contains the location.}
		\label{fig:cellConstraints_2}
	\end{subfigure}%
	\begin{subfigure}{\subfigurewidth+1em}
		\includegraphics[width=\subfigurewidth]{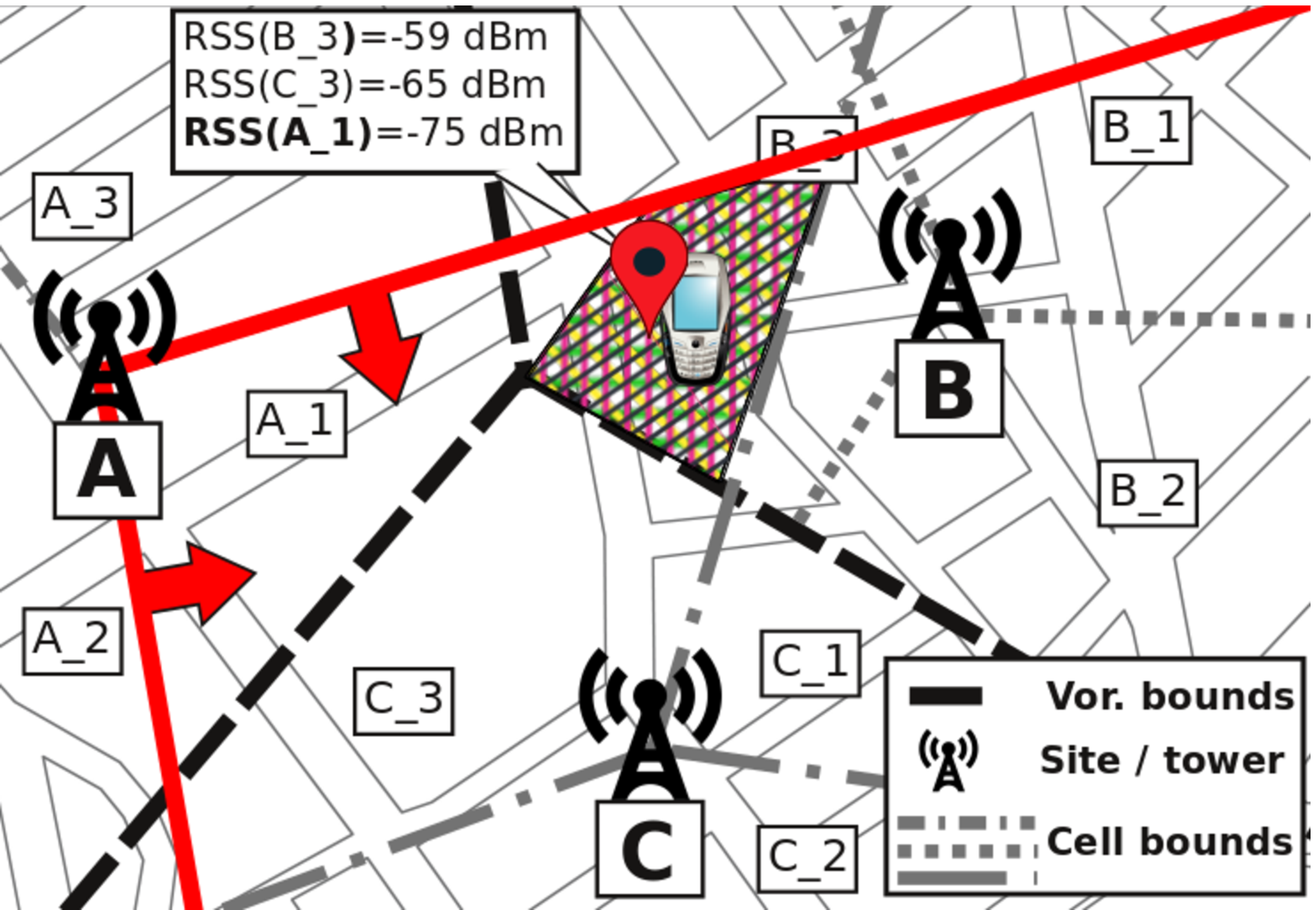}%
		\hfill
		\caption{Cell $A\_1$ contains the location.}
		\label{fig:cellConstraints_3}
	\end{subfigure}%
	\caption{Step 2: Applying cell constraints. Sequential intersections of visible cell sectors' areas further reduce the ambiguity region.}
	\label{fig:cellConstraints}
\end{figure*}

\begin{figure}[!t]
	\centering
	\includegraphics[width=0.8\columnwidth]{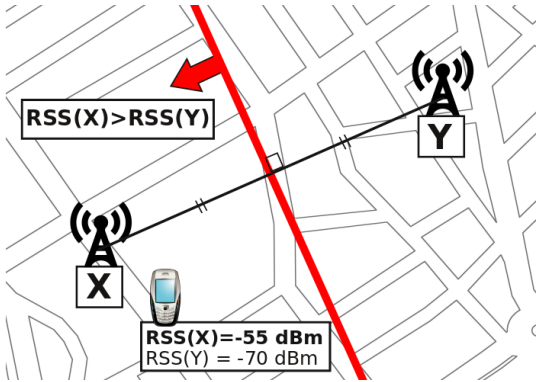}
	\caption{For any pair of sites: if the RSS from site $X$ at the device is greater
		than the RSS from site $Y$, then the device must be closer to $X$ than
		$Y$. This maps to placing the device in the half-plane defined
		by the bisector line between $X$ and $Y$ and containing site $X$.}
	\label{fig:halfPlane}
\end{figure}

\subsubsection{Incremental Voronoi Tessellation}
	\sys{} starts by building the Voronoi diagram~\cite{aurenhammer1991voronoi} of the area of interest with sites as seeds as shown in the example in Fig.~\ref{fig:pairwiseConstraints_1}, where the Voronoi diagram is built for the three towers $A$, $B$, and $C$. A Voronoi diagram splits the area into $n$ non-overlapping polygons, where $n$ is the number of sites/seeds. Each Voronoi polygon is associated with a site/seed $s$ and contains all points in the area that are closer to $s$ than to any other site/seed.
	Since the general trend for RSS is to increase as distance decreases and vice versa, \sys{} initially places the user in the Voronoi polygon of the site with the strongest signal, which represents the points that are closest to this site compared to any other site. 
	In our example cell $B\_3$ is the strongest cell with $-59dBm$, and hence the Voronoi polygon of site $B$ is the initial ambiguity area as shown in Fig.~\ref{fig:pairwiseConstraints_2}. However, this initial area can be very large, especially for a sparse cellular network where the sites/seeds are separated by larger distances. 
	To refine this initial area the first step is to use \textit{Pairwise Site/Cell Constraints}. For every pair of \textit{secondary} sites, i.e. sites other than the strongest site, the area of interest can be split into two half-planes as shown in Fig.~\ref{fig:halfPlane}. Points in the half-plane of site $X$ are closer to site $X$ or receive a higher RSS from cells located at site $X$ as compared to site $Y$, and vice versa. This relation can be leveraged to reduce the ambiguity area. First of all, for every pair of visible sites/cells,
	their corresponding RSS can be used to determine in which half-planes the MU is situated. In total $o({v\choose 2})$ pairwise comparisons 
	indicate in which half-planes the MU is located, where $v$ is the number of visible sites/cells. The intersections between the initial ambiguity area (the Voronoi polygon of the strongest site) and the half-planes resulting from pairwise comparisons of other different sites reduces the ambiguity area significantly. In our example $v=3$, and only one half-plane resulting from the comparison between sites $C$ and $A$ can be used to refine the area as indicated by the arrow in Fig.~\ref{fig:pairwiseConstraints_3}. Specifically, since the RSS from cell $C$ is stronger than the RSS from cell $A$, the MU is placed in the half-plane of site $C$. Intersection with the initial Voronoi polygon results in the reduced area highlighted in Fig.~\ref{fig:pairwiseConstraints_3}.
	\label{sub:vor_constraints}
	\subsubsection{Leveraging Cell Information}
	\label{sub:cell_info}
	The second step to further refine the remaining ambiguity area incorporates cell and sector information. Since a cell is only visible within its sector, the MU should be contained within sectors of all visible cells. Therefore, by calculating the intersection area between the remaining ambiguity area and sectors of all visible cells, the ambiguity area can be further reduced significantly. This is illustrated in Fig.~\ref{fig:cellConstraints}, where intersection calculation between the ambiguity area after applying Pairwise Site Constraints, and sector areas of the three visible cells $B\_3$, $C\_3$ and $A\_1$ results in the reduced area shaded in Fig.~\ref{fig:cellConstraints_3}, whose center of mass is deemed the location estimate. 
	\subsubsection{Discussion}
	Note that \sys{} \textit{does not depend on any propagation model} but leverages only the \textbf{relative RSS} information between each pair of cells as well as the sectorization to obtain a tight ambiguity area for the user's location. 
	
	The only information required by \sys{} to function is the location of the different cell towers and the sectorization information. This is readily available if the system is implemented at the provider side or can be calculated from crowdsourced data \cite{opencellid,li2017identifying}.
	
	Area calculations are computationally expensive and might adversely affect the real-time response of the system. In the next section, we provide the details of the \sys{} system, relaxing the ideal assumptions and handling the computational efficiency aspects.

\section{The \sys{} System}\label{sec:full}
In this section, we provide the \sys{} system architecture followed by the details of its modules that provide a calibration/infrastructure-free and ubiquitous localization system. We assume a network structure similar to that presented in Section~\ref{sub:network_structure} with the typical noisy RF propagation characteristics, therefore relaxing the ideal assumptions of the previous section. Specifically, the heard cells by the MU might be located at the same site or covering the same sector. Additionally, due to noise, cells might be visible outside their ideal sector area. We also describe how \sys{} can compute the area intersections efficiently.

\begin{figure}[!t]
	\centering
	\includegraphics[width=1\columnwidth]{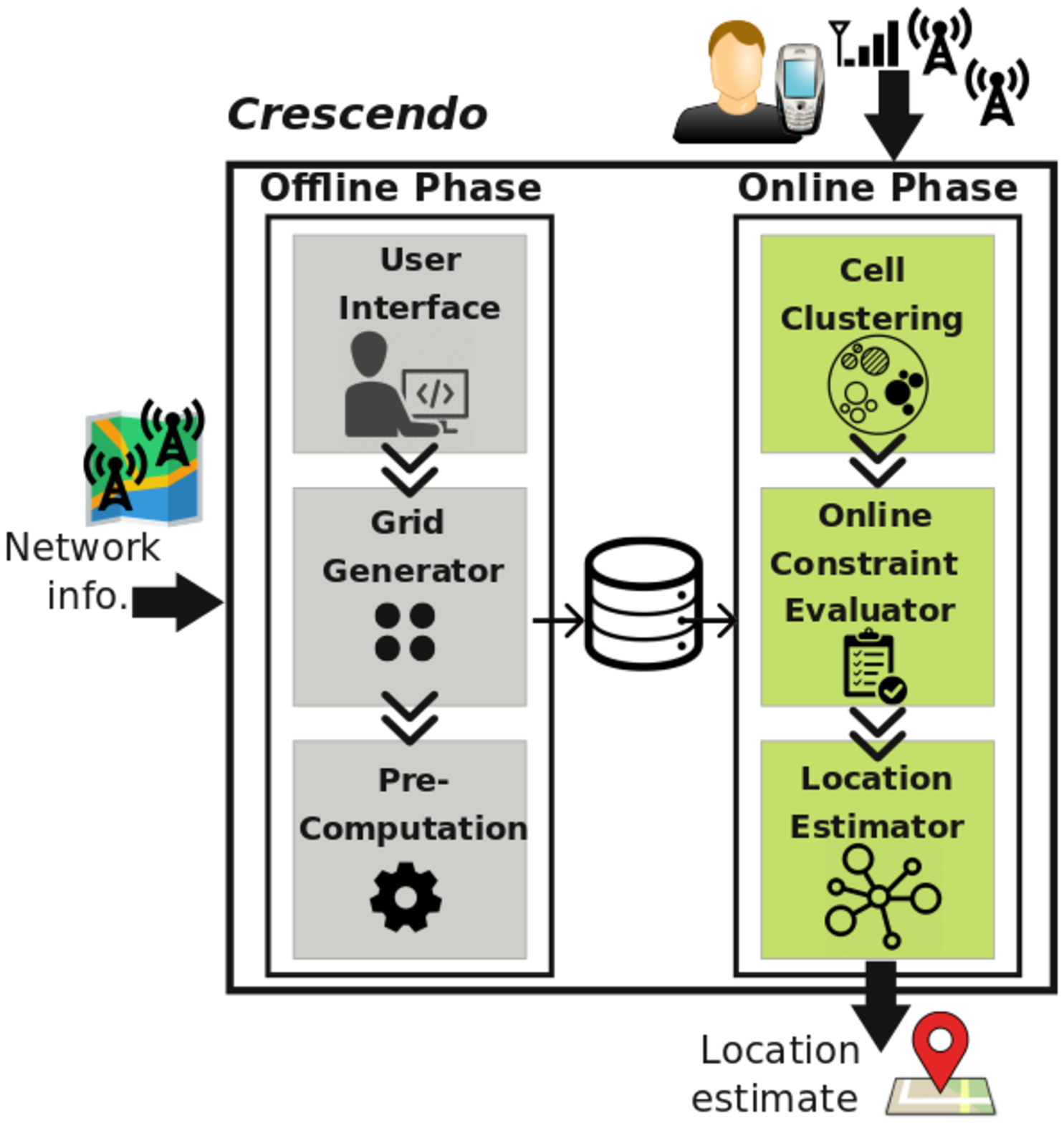}
	\caption{\sys{} system architecture.}
	\label{fig:architecture}
\end{figure}

\subsection{Overview}
Fig.~\ref{fig:architecture} shows the system architecture. The system works in two phases: an offline phase and an online tracking phase.
\subsubsection{Offline Phase}
During the offline phase, the system administrator uses the \textit{\textbf{User Interface}} module to import/enter the required network information (e.g. site IDs and locations, cell and sector information, etc.) A \textit{virtual} grid is  generated and superimposed over the area of interest during this phase using the \textit{\textbf{Grid Generator}} module. This virtual grid is used to speed up calculations as well as handling the inherent noise in RF propagation. The \textit{\textbf{Pre-Computation}} module finally calculates the ``discrete'' Voronoi diagram of the area of interest, which is used to determine the initial user ambiguity area. It also pre-calculates associated parameters with each grid point (e.g. Pairwise Site Constraints and Containing Cell Set). This information is used during the online phase by the \textit{\textbf{Online Constraint Evaluator}} module to reduce the running time and handle noise. 

\subsubsection{Online Tracking Phase}
This is the main operational phase of \sys{}, where location estimates are generated based on the visible network cells. The \textit{\textbf{Cell Clustering}} module starts by clustering visible cells based on their tower locations and using the strongest cell from each tower to represent this tower. In the \textit{\textbf{Online Constraint Evaluator}} module, representative site RSS values obtained from the \textit{\textbf{Cell Clustering}} module are used to create online \textit{Pairwise Site Constraints}. Additionally, \textit{all visible cells} are included in the online \textit{Containing Cell Set}. Based on these constraints, it also  assigns scores to the different grid points based on their likelihood of being the points where the MU is located. Finally, the \textit{\textbf{Location Estimator}} module uses the grid points scores to estimate the final MU location.

\begin{figure}[!t]
	\centering
	\includegraphics[width=0.8\columnwidth]{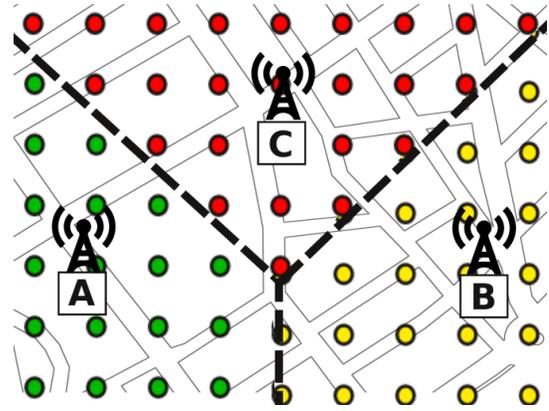}
	\caption{The proposed gridding approach. Grid points representing the ``discrete'' Voronoi polygon of each cell have the same color.}
	\label{fig:grid}
\end{figure}

 \subsection{Virtual Gridding}
 \label{sub:gridding}
 
 Area calculations are computationally expensive and might adversely affect the real-time response of the system. To overcome this issue, we use a grid-based discretization approach. First, a discrete grid with a certain step size (e.g. 50m) is super-imposed on the area of interest as shown in Fig.~\ref{fig:grid}.
 Then, instead of performing area intersection  computations, we compare the \textit{\textbf{expected}} constraints at each location (based on the \textit{\textbf{location}} of the grid point relative to each cell) with the actual constraint state at each location (based on the \textit{\textbf{signal}} heard from the different cells).
 
 In particular, \textit{Offline Pairwise Site Constraints} are calculated for each grid point by comparing the distance of the grid point to each pair of sites and creating a constraint that indicates that e.g. \textit{``Site $A$ is closer than site $B$''} if the distance of the grid point to site $A$ is less than that to site $B$. For $n$ sites a total of $n\choose 2$ constraints is created for each grid point. Additionally, the \textit{Containing Cell Set} is calculated for each grid point. A cell is added to the Containing Cell Set of a grid point if the grid point lies within the area defined by the sector of the cell assuming the sector area extends to infinity. If a cell belongs to the Containing Cell Set of a grid point, then all cells covering the same sector also belong to this containing set. A grid point can naturally be contained in cells from different sites. These distance-based calculations are \textbf{pre-calculated offline} and stored to speed up computations during the online phase. 

While the system is running in real-time, grid points are scanned and the actual constraint value based on the RSS by the user device from the two sites is
 compared to the expected constraint value stored for each grid
 point by the \textit{Online Constraint Evaluator} module. Additionally, visible cells are compared to cells in the Containing Cell Set of the grid point. All grid points start with a score of zero and each matching constraint leads to increasing the grid point's score by 1.
   
\subsection{Location Estimator}
The \textit{\textbf{Location Estimator}} module extracts the grid points with maximum matching scores. Then it calculates the final location estimate as the center of mass of these grid points.

\subsection{Discussion}
Note that the virtual gridding approach allows us to replace the expensive area computations by evaluations at discrete samples.  These evaluations can be calculated offline and stored, significantly speeding up the online phase as we quantify in the evaluation section. This is further enhanced by the clustering performed by the strongest site/tower during the online phase.

In addition to reducing the computational complexity, the virtual gridding also allows us to better handle the noise in the RF signal. For example, in reality, when a cell is visible outside its sector area, this can lead to contradictions leading to an empty intersection when applying the basic idea without modification. The proposed virtual gridding approach solves this problem by assigning a score to each grid point based on its matching constraints. The MU
estimated location is now based on the grid points that match the largest number of actual and expected constraints.

\section{Evaluation}
\label{sec:eval}
In this section, we evaluate the proposed algorithm in two testbeds; an urban area of $0.507 \textrm{km}^2$ with a dense cellular network and a rural area of $0.723 \textrm{km}^2$ with a relatively sparse cellular network; using $22953$ samples. Testbed properties are summarized in Table~\ref{tab:testbeds}. We first start by examining the effect of the different parameters on accuracy.
Then, we compare our proposed algorithm to two of the most commonly used low-overhead outdoor localization algorithms, the Cell ID method and the Centroid method.

\begin{table}[htbp]
	\caption{Testbed Properties}
	\begin{center}
		\begin{tabular}{|c|c|c|}
			\hline
			\textbf{Property}&\multicolumn{2}{|c|}{\textbf{Testbed}} \\
			\cline{2-3} 
			\textbf{} & \textbf{\textit{Urban}}& \textbf{\textit{Rural}}\\
			\hline
			Area ($\textrm{km}^2$)& $0.507$&$0.723$  \\
			\hline
			Cell density (cells/$\textrm{km}^2$) & 224&100   \\
			\hline
			\sys{} median error ($m$)&\textbf{152} &\textbf{224}\\
			\hline
			Cell ID median error ($m$)&187(-23\%) &545(-143\%)\\
			\hline
			Centroid median error ($m$)&214(-40.7\%)&264(-17.8\%)\\
			\hline
		\end{tabular}
		\label{tab:testbeds}
	\end{center}
\end{table}

\begin{table}[htbp]
	\caption{Default parameter values used in evaluation}
	\begin{center}
		\begin{tabular}{|c|c|c|}
			\hline
			\textbf{Parameter}&Range&Default value \\
			\hline
			Grid size (m)& $25$-$200$&$50$  \\
			\hline
			Cell density(cells/$\textrm{km}^2$)& $21$-$224$&$224$  \\
			\hline
			Testbed&Urban, rural&Urban\\
			\hline
		
		\end{tabular}
		\label{tab:params}
	\end{center}
\end{table}

\subsection{Parameters Effect}
We evaluate the effect of grid size and cell density on system performance. Table~\ref{tab:params} shows the default parameter values used in evaluation.

\subsubsection{Grid Size}
\begin{figure}[!t]
	\centering
	\includegraphics[width=1\columnwidth]{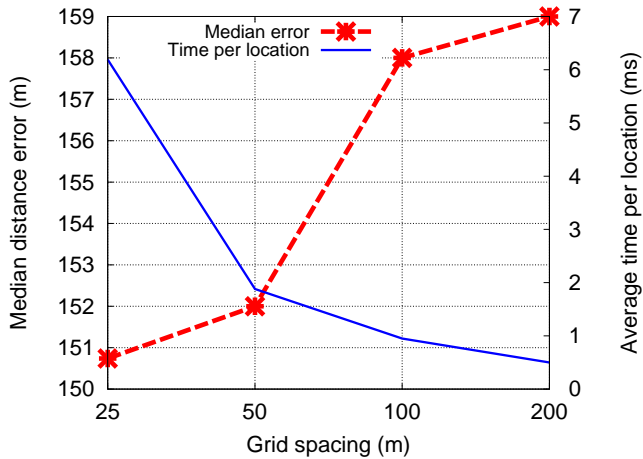}
	\caption{Effect of changing the grid size.}
	\label{fig:grid_effect}
\end{figure}

Fig.~\ref{fig:grid_effect} shows the effect of changing the virtual grid size on system performance. The figure shows that increasing the grid size reduces the accuracy while significantly reducing the time required per location estimate due to the reduced number of virtual grid points.
\subsubsection{Cellular Network Density}
\label{sub:eval_density}
\begin{figure}[!t]
	\centering
	\includegraphics[width=1\columnwidth]{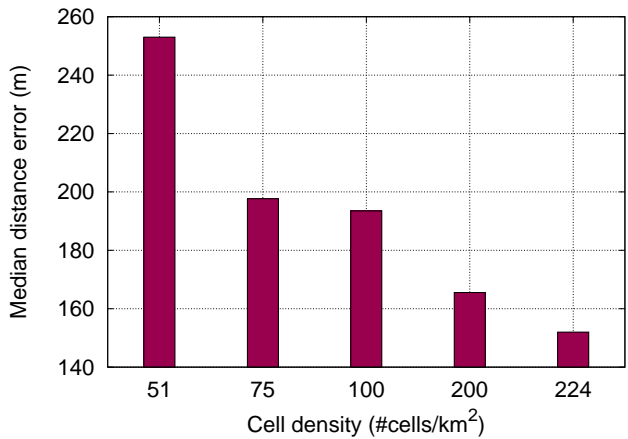}
	\caption{Effect of changing cell density.}
	\label{fig:density}
\end{figure}

Fig.~\ref{fig:density} shows the effect of changing the cells density. This was obtained by uniformly dropping cells. The figure shows that, as expected, reducing the cells density leads to reducing the system's accuracy due to the reduction of the available information. Nonetheless, using the default cell density, \sys{} can achieve accuracy of 152m. This also explains the higher accuracy of the system in urban areas compared to rural ones.

\subsection{Comparison to Other Systems}
\label{sub:eval_comp}
\begin{figure*}[htp]
	\setlength{\subfigurewidth}{.45\textwidth}
	\begin{subfigure}{\subfigurewidth+1em}
		\includegraphics[width=\subfigurewidth]{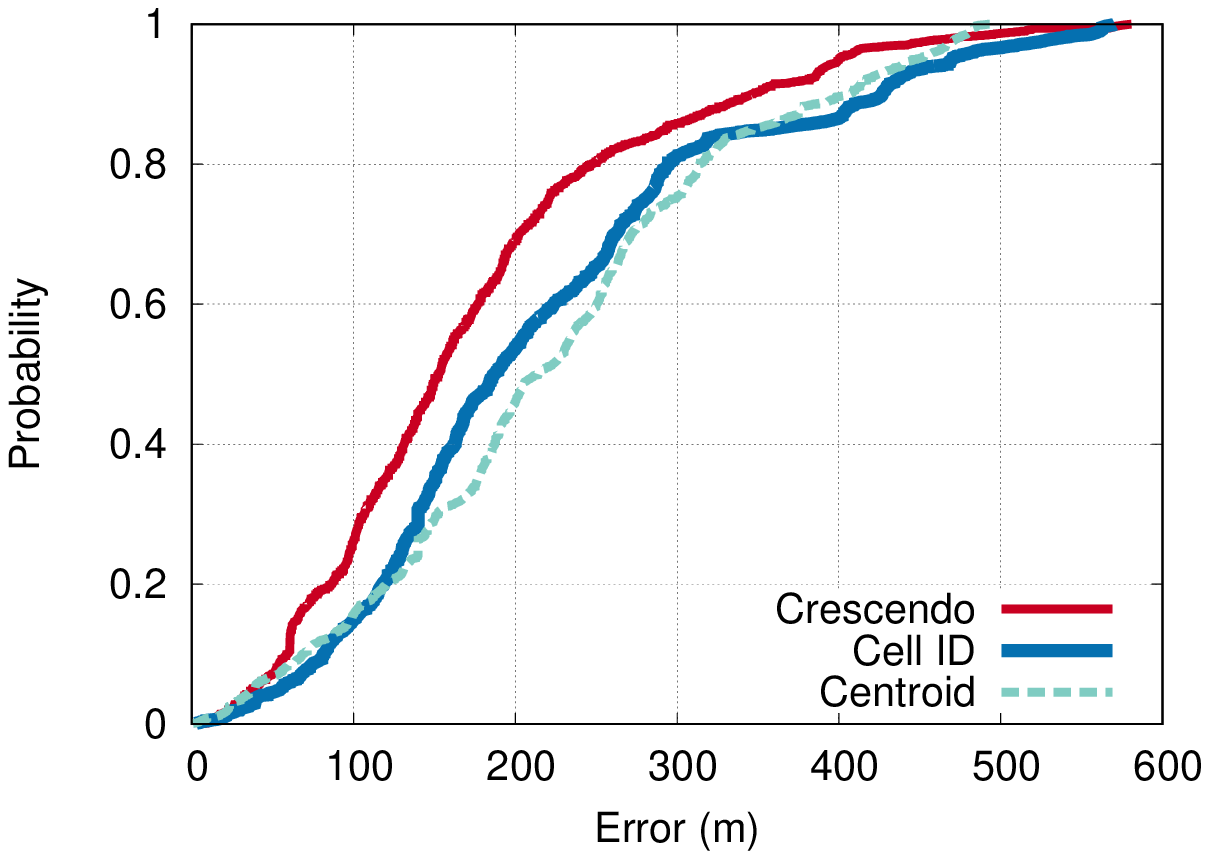}%
		\hfill
		\caption{Urban area.}
		\label{fig:comp_urban}
	\end{subfigure}\hspace{5mm}%
	\begin{subfigure}{\subfigurewidth+1em}
		\includegraphics[width=\subfigurewidth]{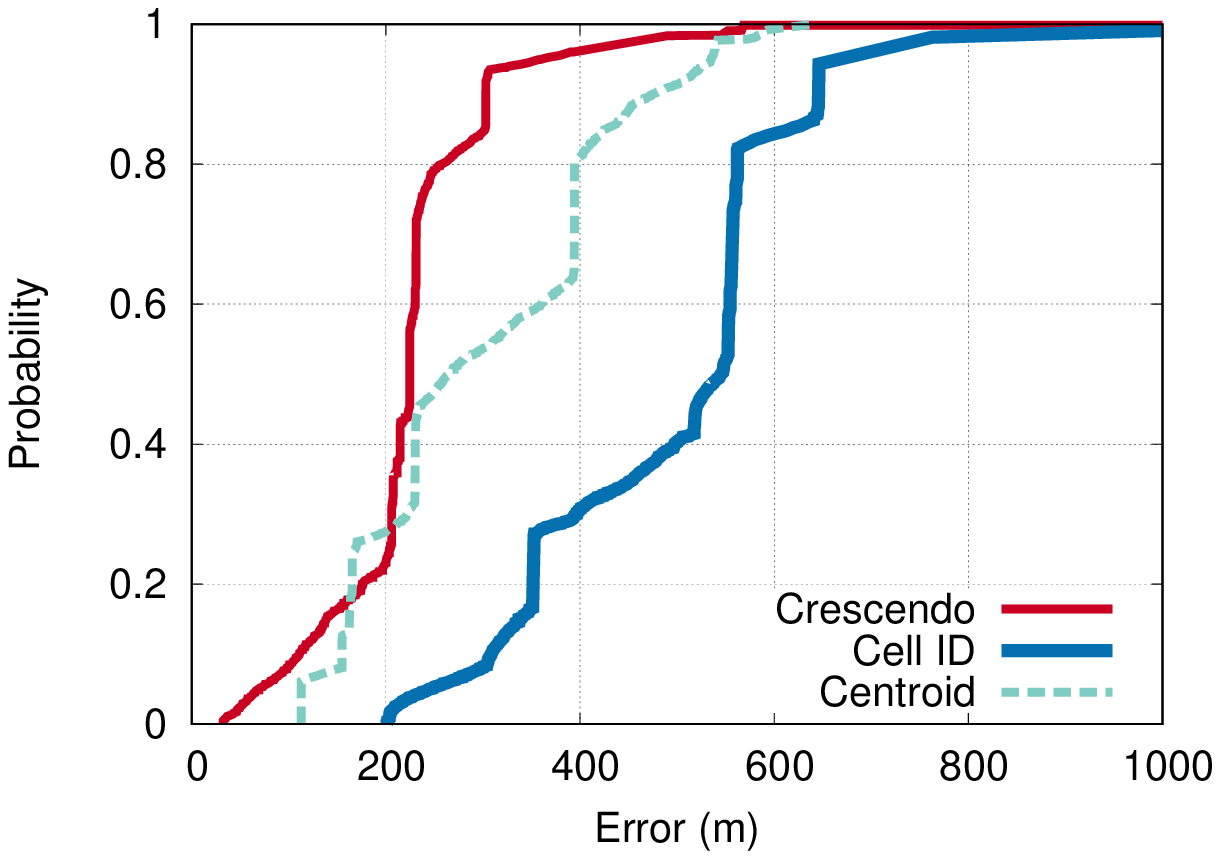}%
		\hfill
		\caption{Rural area.}
		\label{fig:comp_rural}
	\end{subfigure}%
	\caption{Comparison to other techniques.}
	\label{fig:eval_comparison}
\end{figure*}

We compare our proposed technique to two of the most commonly used infrastructure-free techniques: the Cell ID~\cite{dufkova2008active} and the Centroid method. In the Cell ID method, the location of the MU is estimated as the location of the strongest visible cell while in the Centroid method, the location is estimated as the center of mass of all visible sites. As shown in Fig.~\ref{fig:eval_comparison}, \sys{} outperforms both methods by at least 18\% and 15\% in the urban and rural area, respectively. This is due to its novel incremental Voronoi tessellation technique and the use of sector information in addition to RSS.

\section{Conclusion}
\label{sec:conc}
In this paper, we presented \sys{}, a ubiquitous low-overhead outdoor localization technique that works for all cell phones with no calibration and no additional infrastructure support. \sys{} depends on the Voronoi diagram of network sites, pairwise comparison between sites, as well as cell sector information to incrementally improve the localization accuracy without the need for data collection or special sensors.

We implemented and tested \sys{} in an urban and a rural area and compared its performance against existing techniques. Results show median accuracies of 152m and 224m in the urban and rural area, respectively, and an improvement over classical techniques of at least 18\% and 15\%, respectively.

\section*{Acknowledgment}
This work has been supported by a grant from the
Egyptian National Telecommunications Regulatory Authority
(NTRA).

\bibliographystyle{IEEEtran}

\bibliography{incVorOut_IEEEfull_short}

\end{document}